\documentclass[aps,prl,twocolumn,groupedaddress]{revtex4}

\usepackage{graphicx}
\usepackage{bm}
\usepackage{bbm}
\usepackage{amsmath}
\usepackage{subfigure}

\begin{document}

\title{Theory of Polar Blue Phases}
\author{Shaikh M. Shamid}
\author{David W. Allender}
\author{Jonathan V. Selinger}
\email{jselinge@kent.edu}
\affiliation{Liquid Crystal Institute, Kent State University, Kent, OH 44242}

\date{May 19, 2014}

\begin{abstract}
In liquid crystals, if flexoelectric couplings between polar order and director gradients are strong enough, the uniform nematic phase can become unstable to formation of a modulated polar phase.  Previous theories have predicted two types of modulation, twist-bend and splay-bend; the twist-bend phase has been found in recent experiments.  Here, we investigate other types of modulation, using lattice simulations and Landau theory.  In addition to twist-bend and splay-bend, we also find polar blue phases, with 2D or 3D modulations of both director and polar order.  We compare polar blue phases with chiral blue phases, and discuss opportunities for observing them experimentally.
\end{abstract}

\maketitle

Nematic liquid crystals exhibit flexoelectric couplings between polar order and gradients in the director field.  These couplings lead to the flexoelectric effect, in which bend or splay of the director induces a net electrostatic polarization, and the converse flexoelectric effect, in which an applied electric field induces bend and splay~\cite{meyer69}.  When the couplings become strong enough, the uniform nematic phase can become unstable to the formation of a modulated polar phase.  The question is then:  What is the structure of the modulated polar phase?  Classic work by Meyer~\cite{meyer76} and further studies by other investigators~\cite{dozov2001,memmer2002,shamid13} predicted two possible structures, known as twist-bend and splay-bend.  One of these predictions, the twist-bend phase, has recently been identified in experiments on bimesogens~\cite{panov2010,luckhurst2011,noelclark13,oleg13}, and it is now recognized as a remarkable experimental discovery of a liquid-crystal phase with a new type of symmetry.

The purpose of this paper is to investigate whether liquid crystals can form modulated polar phases with other structures, different from twist-bend and splay-bend.  This study is motivated by an analogy between polarity and chirality.  The twist-bend phase of polar liquid crystals is similar to the cholesteric phase of chiral liquid crystals; both are helical phases induced by molecular assymetry.  Apart from the cholesteric phase, chiral liquid crystals also form blue phases, with complex three-dimensional (3D) modulations of the director~\cite{wright89}.  We would like to consider whether polar liquid crystals can form analogous blue phases.  To explore this question, we use both lattice simulations and Landau theory.  For lattice simulations, we use generalized Lebwohl-Lasher models, which we constructed in previous studies of splay~\cite{subas10} and bend~\cite{shamid13} flexoelectricity.  For Landau theory, we generalize earlier theories of chiral blue phases, which represent the liquid-crystal order tensor as a series of plane waves and minimize the free energy over the Fourier coefficients~\cite{brazovskii1975,hornreich1983}.

Both of our approaches demonstrate that the system can form polar blue phases.  One polar blue phase is a 3D body-centered-cubic (bcc) lattice, resembling a lattice of micelles, which is dominated by splay in the director.  Another such phase is a 2D hexagonal lattice, dominated by bend in the director.  These results can be compared with experimental studies of complex modulated phases in supramolecular liquid crystals~\cite{ungar03}.

Our results can be contrasted with other theories of modulated phases in liquid crystals.  In the theory of Alexander and Yeomans~\cite{alexander2007}, an applied electric field can induce a flexoelectric blue phase, with uniform polar order and modulated director.  Our work considers polar order that forms spontaneously, not induced by a field, and hence the system forms a more complex modulation in the polar order as well as the director.  In the theory of Castles et al.~\cite{castles2010}, flexoelectric couplings help to stabilize chiral blue phases, thus increasing the blue-phase temperature range in chiral liquid crystals.  Our work finds a new type of blue phase even without chirality.  In the theory of Hinshaw et al.~\cite{hinshaw88}, the director and polar order are both modulated within the layers of ferroelectric smectic liquid crystals.  We find similar modulations in nematic phases, without smectic layers.  

\begin{figure}
(a)~\subfigure{\includegraphics[width=2.75in]{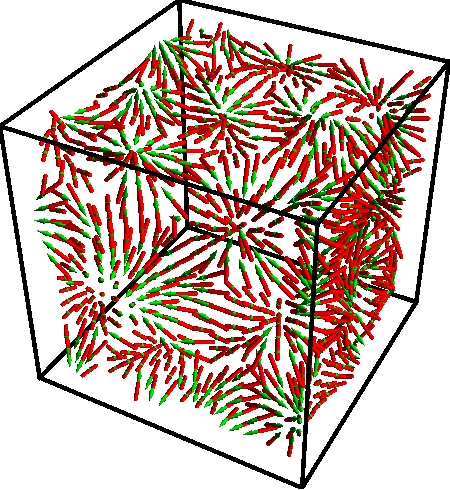}}
(b)~\subfigure{\includegraphics[width=2.75in]{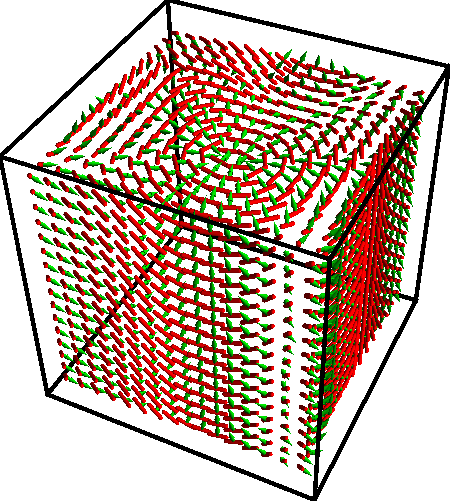}}
\caption{\label{Fig1} (Color online) Equilibrium configurations from Monte Carlo simulations.  (a)~Splay model with parameters $A = 1.25$, $B = 0.01$, $C = 3.0$, and temperature $T = 0.35$, simulated on a $16 \times16 \times 16$ lattice with periodic boundary conditions.  The resulting structure is approximately bcc.  (b)~Bend model with $A = 2.5$, $B_{1} = 0.4$, $B_{2} = 0.5$, $C = -3.0$, $T = 0.45$, simulated on a $16 \times 16 \times 8$ lattice (shown double using periodic boundary conditions, for comparison with previous figure).  The resulting structure is approximately hexagonal.  On each lattice site, the red rod represents the director of the melecules, and the green arrow represents the polar order.  For clarity, only the sites on the surfaces are shown.}
\end{figure}

In simulations, we generalize the classic Lebwohl-Lasher model of nematic liquid crystals~\cite{lebwohl72}, where each lattice site has a spin representing the orientation of the director, obeying the nematic symmetry $\hat{\bm{n}}_i \rightarrow -\hat{\bm{n}}_i$.  To construct a lattice Hamiltonian for splay flexoelectricity, applicable for pear-shaped molecules, we relax that nematic symmetry condition to capture the longitudinal polarity of the melecules.  As shown in our previous study of splay flexoelectricity~\cite{subas10}, the total lattice Hamiltonian can be written as
\begin{eqnarray}
H&=&-\sum_{\langle i,j\rangle}\biggl[
A(\hat{\bm{n}}_i\cdot\hat{\bm{n}}_j)^2 + B(\hat{\bm{n}}_i\cdot\hat{\bm{n}}_j)\nonumber\\
&&\qquad
+C\Bigr(\frac{1+\hat{\bm{n}}_i \cdot\hat{\bm{n}}_j}{2} \Bigr)^2\hat{\bm{r}}_{ij}\cdot(\hat{\bm{n}}_j - \hat{\bm{n}}_i)\Bigr].
\label{hamiltonian}
\end{eqnarray}
Here, the $A$ term favors nematic order, the $B$ term favors polar order, and the $C$ term represents the interaction between splay and polar order.  Our previous study showed that this model has three phases, isotropic, nematic, and polar, but did not examine the structure of the polar phase.  Now, we run Monte Carlo simulations in the low-temperature polar phase to determine the equilibrium structure.  These simulations show that the structure is a lattice of hedgehog defects, or micelles, as shown in Fig.~1(a).  As the flexoelectric coupling $C$ increases, the lattice constant decreases.  We cannot determine the structure precisely because the periodic boundary conditions interfere with the natural periodicity, but the symmetry is approximately bcc. 

Similarly, the lattice Hamiltonian for bend flexoelectricity can be written as~\cite{shamid13}
\begin{eqnarray}
H&=&-\sum_{\langle i,j\rangle}\biggl[
A(\hat{\bm{n}}_i\cdot\hat{\bm{n}}_j)^2
+B_1\hat{\bm{b}}_i\cdot\hat{\bm{b}}_j
+B_2(\hat{\bm{b}}_i\cdot\hat{\bm{b}}_j)^2\nonumber\\
&&\qquad
-\frac{C}{4}\Bigl[(\hat{\bm{b}}_j\cdot\hat{\bm{n}}_i)[\hat{\bm{r}}_{ij}\cdot\{\hat{\bm{n}}_i
+\hat{\bm{n}}_j(\hat{\bm{n}}_i\cdot\hat{\bm{n}}_j)\}] \\
&&\qquad\qquad
-(\hat{\bm{b}}_i\cdot\hat{\bm{n}}_j)[\hat{\bm{r}}_{ij}\cdot\{\hat{\bm{n}}_j
+\hat{\bm{n}}_i(\hat{\bm{n}}_i\cdot\hat{\bm{n}}_j)\}]\Bigr]\biggr],\nonumber
\end{eqnarray}
where $\hat{\bm{n}}_i$ represents the long molecular axis and $\hat{\bm{b}}_i$ the transverse molecular dipole at site $i$.  In this Hamiltonian, the $A$ term favors nematic order of the long axes, the $B_{1}$ term favors polar order of the transverse dipoles,  the $B_{2}$ term favors nematic order of the transverse dipoles, and the $C$ term represents the interaction between bend and polar order.  Our previous study showed that this model has several phases:  isotropic, uniaxial nematic, biaxial nematic, twist-bend, and splay-bend.  We now run further Monte Carlo simulations in the low-temperature polar regime.  In addition to the twist-bend and splay-bend phases with 1D modulation, we also find a new phase with 2D modulation.  This phase consists of a lattice of vortices, as shown in Fig.~1(b).  Each vortex involves bend of the long axis and splay of the polar order.  At each antivortex, the magnitude of the polar order is reduced.  This structure is similar to the prediction for modulated smectic layers by Hinshaw et al.~\cite{hinshaw88}, and to the polarization-modulated smectic phases found experimentally by Coleman et al.~\cite{coleman2003}, but without smectic order.  Again, we cannot determine the structure precisely because of the periodic boundary conditions, but it is approximately a 2D hexagonal lattice.  We do not see any indication of a 3D modulated phase in this model.

To understand polar blue phases further, we develop an analytic Landau theory.  In this theory, we follow the approach developed for chiral blue phases in the ``high chirality'' limit~\cite{wright89,brazovskii1975,hornreich1983}.  For chiral blue phases, those theories expand the nematic order tensor field $Q_{\alpha\beta}(\bm{r})$ as a series of plane waves, with wavevectors corresponding to reciprocal lattice vectors of the proposed lattice.  They then insert the tensor field into a free energy functional that represents the chirality of the phase, and minimize over the Fourier coefficients to determine the optimal structure.  To describe polar rather than chiral blue phases, we must modify the theory in two ways.  First, we must expand \emph{both} the nematic order tensor field $Q_{\alpha\beta}(\bm{r})$ and the polarization vector field $P_{\alpha}(\bm{r})$ as series of plane waves, using appropriate spherical-harmonic modes.  Second, we must use a free energy that represents the coupling between polarity and director gradients, without chirality.

In general, the nematic order tensor field $Q_{\alpha\beta}(\bm{r})$ can be expressed in terms of five modes.  In standard spherical-harmonic notation, these modes are $l=2$ and $m=0,$ $\pm1$, or $\pm2$.  The $m=\pm2$ modes represent right- and left-handed cholesteric helices; these modes are used in the theory of chiral blue phases because they are the most chiral.  The $m=\pm1$ modes represent right- and left-handed tilted conical helices; they can be used to describe the twist-bend and splay-bend phases in terms of $Q_{\alpha\beta}(\bm{r})$ (although previous theories have described these phases in terms of the director rather than the tensor field).  The $m=0$ mode represents a modulation of the nematic order parameter; we will use that mode to describe polar blue phases.  Hence, for a single wavevector $\bm{q}$, we write
\begin{equation}
Q^{\bm{q}}_{\alpha\beta}(\bm{r}) = a_{\bm{q}} \left( \frac{3}{2} \hat{q}_{\alpha} \hat{q}_{\beta} - \frac{1}{2} \delta_{\alpha\beta} \right)
\cos\left( \bm{q} \cdot \bm{r}  + \theta_{\bm{q}} \right).
\end{equation}
Here, $a_{\bm{q}}$ is the amplitude of the mode, $\theta_{\bm{q}}$ is the phase, and $\bm{\hat{q}} = \bm{q}/|\bm{q}|$ is a unit vector.

Similarly, the polarization vector field $\bm{P}(\bm{r})$ can be expressed in terms of three modes, which in spherical-harmonic notation are $l=1$ and $m=0$ or $\pm1$.  The transverse $m=\pm1$ modes represent right- and left-handed helices, which describe polar order in the twist-bend and splay-bend phases.  The longitudinal $m=0$ mode represents a modulation in the magnitude of polar order, and we will use it for polar blue phases.  For a single wavevector $\bm{q}$ we write
\begin{equation}
\bm{P}^{\bm{q}}(\bm{r}) = b_{\bm{q}} \bm{\hat{q}} \cos\left( \bm{q} \cdot \bm{r}  + \phi_{\bm{q}} \right),
\end{equation}
where $b_{\bm{q}}$ is the amplitude and $\phi_{\bm{q}}$ the phase of the mode.

To describe a polar blue phase with any particular symmetry, we must sum both $Q_{\alpha\beta}(\bm{r})$ and $\bm{P}(\bm{r})$ over wavevectors $\bm{q}$ in the reciprocal lattice,
\begin{equation}
Q_{\alpha\beta}(\bm{r}) = \sum_{\bm{q}} Q^{\bm{q}}_{\alpha\beta}(\bm{r}), \quad  \bm{P}(\bm{r}) = \sum_{\bm{q}} \bm{P}^{\bm{q}}(\bm{r}),
\end{equation}
using at least all the reciprocal lattice vectors with the smallest magnitude.  For bcc symmetry, we sum over six wavevectors in the fcc reciprocal lattice,
\begin{equation}
\bm{q} = \left(\frac{q}{\sqrt{2}}, \pm\frac{q}{\sqrt{2}}, 0 \right), \left(0, \frac{q}{\sqrt{2}}, \pm\frac{q}{\sqrt{2}} \right),  
\left(\pm\frac{q}{\sqrt{2}}, 0, \frac{q}{\sqrt{2}} \right).
\end{equation}
Each of these modes must have the same amplitude $a$, but they may have different phases $\theta_{\bm{q}}$ and $\phi_{\bm{q}}$.  Three linear combinations of the phases are arbitrary, corresponding to uniform translations in $x$, $y$, or $z$.  All other linear combinations affect the structure, and must be determined by minimizing the free energy.

Likewise, for 2D hexagonal symmetry, we sum over three wavevectors in the hexagonal reciprocal lattice,
\begin{equation}
\bm{q} = \left(q,0,0 \right), \left(-\frac{q}{2}, \pm\frac{\sqrt{3} q}{2},0 \right).
\end{equation}
Here, two linear combinations of the phases are arbitrary, corresponding to uniform translations in $x$ or $y$.  All other linear combinations affect the structure, and must be determined by minimizing the free energy.

We write the general free energy density for a liquid crystal of polar molecules as
\begin{eqnarray}
\label{fcomplete}
F&=& \frac{1}{2} A \left(Q_{\alpha \beta}Q_{\alpha \beta} \right)+\frac{1}{3}B \left(Q_{\alpha \beta}Q_{\beta \gamma}Q_{\gamma \alpha} \right) \\
&& + \frac{1}{4}C\left(Q_{\alpha \beta}Q_{\alpha \beta} \right)^2 +\frac{1}{2} L_{1} \left( \partial_{\alpha}Q_{\beta \gamma}\partial_{\alpha}Q_{\beta \gamma} \right) \nonumber\\
&& + \frac{1}{2} L_{2} \left( \partial_{\alpha}Q_{\alpha \gamma}\partial_{\beta}Q_{\beta \gamma} \right) + \frac{1}{2}\mu \left|\bm{P}\right|^2
+\frac{1}{4}\nu\left|\bm{P}\right|^4\nonumber\\
&& + \frac{1}{2}\kappa \left( \bm{\nabla}\bm{P} \right)^2 - \lambda  P_{\alpha} \partial_{\beta}Q_{\alpha \beta} + \eta  P_{\alpha} Q_{\alpha \beta} P_{\beta}. \nonumber
\end{eqnarray}
This expression is similar to the free energy in our previous paper, but expressed in terms of the nematic order tensor $Q_{\alpha\beta}$ instead of the director.  Here, the $A$, $B$, and $C$ terms are the standard Landau-de Gennes free energy expansion in powers of $Q_{\alpha\beta}$.  The $L_1$ and $L_2$ terms are the lowest-order terms in the Frank free energy, expressed in terms of $Q_{\alpha\beta}$.  The $\mu$, $\nu$, and $\kappa$ terms are the Ginzburg-Landau expansion in powers of the polar order parameter $\bm{P}$. The $\lambda$ term is the flexoelectric coupling between $\bm{P}$ and gradients in $Q_{\alpha\beta}$, which may be either splay or bend.  Finally, the $\eta$ term favors alignment of $\bm{P}$ with respect to $Q_{\alpha\beta}$.  If $\eta<0$, the favored polarization is parallel to the director, leading to splay flexoelectricity.  If $\eta>0$, the favored polarization is perpendicular to the director, leading to bend flexoelectricity.

\begin{figure}
(a)~\subfigure{\includegraphics[width=2.75in]{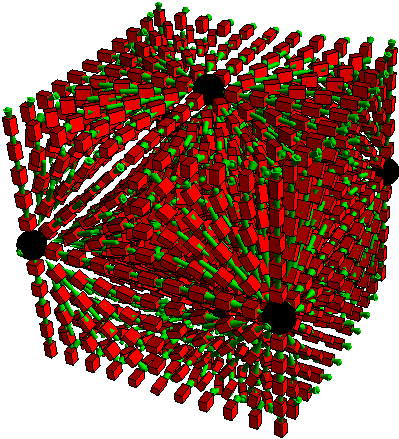}}
(b)~\subfigure{\includegraphics[width=2.75in]{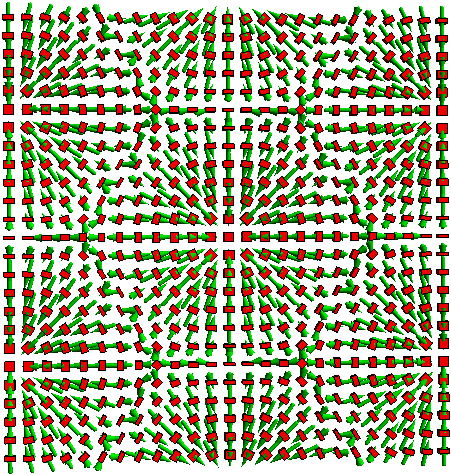}}
\caption{\label{Fig2} (Color online) Visualization of polar blue phases obtained from Landau theory.  (a) 3D bcc lattice, with black dots indicating the centers of the hedgehogs.  (b) 2D hexagonal structure, in a cross section through the $(x,y)$ plane.  In each case, the nematic order is represented by red boxes corresponding to the three eigenvalues and eigenvectors of $Q_{\alpha\beta}(\bm{r})$, and the polar order is represented by green arrows with the magnitude and direction of $\bm{P}(\bm{r})$.}
\end{figure}

We can now determine the free energies of the bcc and hexagonal structures.  For each structure, we insert the Fourier series expressions for $Q_{\alpha\beta}(\bm{r})$ and $\bm{P}(\bm{r})$ into the free energy density of Eq.~(\ref{fcomplete}), and average over a full unit cell, to obtain the free energy as a function of amplitudes $a$ and $b$, phases $\theta_{\bm{q}}$ and $\phi_{\bm{q}}$, and the wavevector magnitude $q$.  First, we minimize over the phases, which enter only into the cubic and quartic terms of the free energy.  This minimization gives the bcc and hexagonal structures shown in Figs.~2(a) and~2(b).  Note that these structures are similar to the simulation results of Figs.~1(a) and~1(b).  The bcc structure is a 3D lattice of hedgehogs, with splay in both $\bm{P}(\bm{r})$ and the main eigenvector of $Q_{\alpha\beta}(\bm{r})$.  The hexagonal structure is a 2D lattice with splay in $\bm{P}(\bm{r})$ and bend in the main eigenvector of $Q_{\alpha\beta}(\bm{r})$.  In both cases, $Q_{\alpha\beta}(\bm{r})$ is highly biaxial in much of the structure, as should be expected when it is represented by a small number of Fourier modes.  This biaxiality might be suppressed if we used more Fourier modes (higher-order reciprocal lattice vectors), or if we performed a numerical minimization in real space.

\begin{figure}
\includegraphics[width=3.375in]{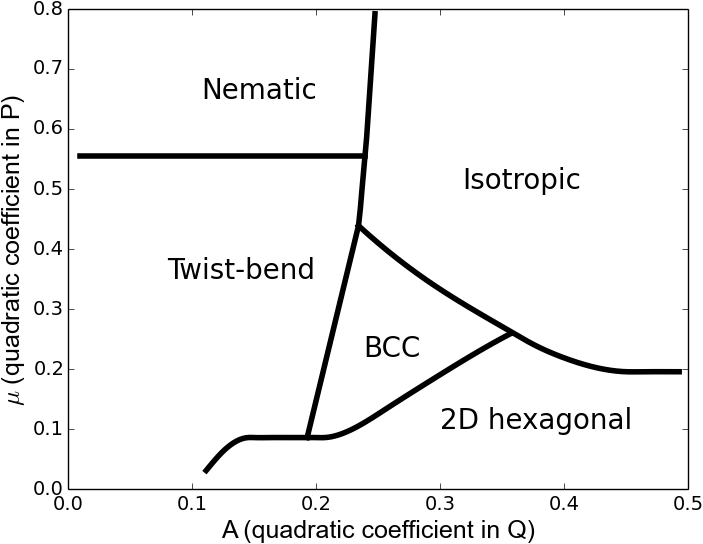}
\caption{\label{Fig3} Numerical phase diagram in terms of the two quadratic coefficients $A$  and $\mu$, which are expected to vary with temperature, for fixed parameters $\lambda = 1.5$, $B = -2.4$, $\eta = -0.18$, $C = L_{1} = L_{2} = \kappa = \nu = 1 $.  For these parameters, all transitions are first-order.}
\end{figure}

Next, we minimize the free energy of each structure, bcc and hexagonal, over the variational parameters $a$, $b$, and $q$.  We then compare the free energies of these two structures, as well as the isotropic phase (with $Q_{\alpha\beta}=0$ and $\bm{P}=0$), nematic phase (with uniform $Q_{\alpha\beta}\not=0$ and $\bm{P}=0$), twist-bend and splay-bend phases (described with $m=\pm1$ spherical harmonic modes for $Q_{\alpha\beta}(\bm{r})$ and $\bm{P}(\bm{r})$), and 3D simple cubic and 2D square lattices (described with $m=0$ spherical harmonic modes analogous to bcc and hexagonal).  We obtain the numerical phase diagram shown in Fig.~3.  This phase diagram is expressed in terms of the two quadratic coefficients $A$ and $\mu$ in the free energy, which are both expected to vary with temperature; the other model parameters are held fixed ($B$, $C$, $L_{1}$, $L_{2}$, $\nu$, $\kappa$, $\lambda$, and $\eta$).  Reducing $A$ favors nematic order, while reducing $\mu$ favors polar order.  Hence, the lower left of the phase diagram represents low temperature, and the upper right represents high temperature.

The phase diagram shows regions of the isotropic, nematic, twist-bend, bcc, and 2D hexagonal phases.  Each of these phases is stable for some set of numerical parameters.  We have not found any parameters for which the splay-bend, simple cubic, or 2D square phases are stable.  Indeed, we would not expect the simple cubic or 2D square phases to be stable in this model, because the free energy includes cubic terms ($B Q_{\alpha \beta}Q_{\beta \gamma}Q_{\gamma \alpha}$ and $\eta  P_{\alpha} Q_{\alpha \beta} P_{\beta}$).  These terms can be negative for the bcc and 2D hexagonal structures, but they are zero for the simple cubic and 2D square structures (assuming that only the lowest-order reciprocal lattice vectors contribute to the modulations).  Of course, we recognize that our predicted phase diagram is limited in several ways:  we have only explored a limited range of numerical parameters; the free energy could include higher-order terms; and the system could form modulations with different lattice symmetries or different spherical-harmonic indices $m$ than we have considered.  Nevertheless, the phase diagram gives an indication of the phases and phase sequences that can be found in a simple model.

As noted in the introduction, the twist-bend phase has recently been observed in experiments on bimesogens~\cite{panov2010,luckhurst2011,noelclark13,oleg13}.  The polar blue phases have not yet been reported in bimesogens, although future experiments might look for those phases.  To our knowledge, the best experimental realization of polar blue phases is the work on supramolecular liquid crystals by Ungar et al.~\cite{ungar03}.  These materials are composed of dendrons, with the shapes of cones or flat wedges (like pizza slices).  The cone shape should favor splay flexoelectricity, while the flat wedge shape should favor a combination of splay flexoelectricity and biaxiality.  Indeed, experiments show that these systems form phases with complex 3D modulated structures, including a bcc lattice and a more complex tetragonal structure.  These results have previously been modeled through geometric arguments~\cite{ziherl01} and molecular simulation~\cite{cleaver06}.  Here, we emphasize that these structures can be regarded as polar blue phases, with local nematic and polar order parameters that are modulated in periodic lattices.  These order parameters can be described by plane waves with appropriate spherical-harmonic modes, just as in chiral blue phases.

In conclusion, this paper has explored a general analogy between chirality and polarity in liquid crystals.  Just as chirality induces a spontaneous twist of the director, polarity induces spontaneous splay and bend.  The spontaneous twist of chiral liquid crystals most commonly leads to a cholesteric phase, but it can also lead to more complex blue phases.  Similarly, experiments have already shown that the spontaneous bend of polar liquid crystals leads to a twist-bend phase, and we now argue that it can also lead to polar blue phases.  This argument provides a way to interpret modulated structures in supramolecular liquid crystals, and it predicts that similar structures may form in bimesogens.

We thank D. J. Cleaver for helpful discussions.  This work was supported by NSF Grant DMR-1106014.

\end{document}